\documentclass{raa}
\usepackage{natbib,times}
\usepackage{lscape}
\bibpunct{(}{)}{,}{a}{}{;}

\providecommand{\bm}[1]{\mbox{\boldmath$#1$\unboldmath}}

\begin{document}

\title{Discovery of four gravitational lensing systems by
clusters in the SDSS DR6
 $^*$
\footnotetext{\small $*$ Supported by the National Natural Science
Foundation of China.} }

\volnopage{{\bf 2009} Vol.\ {\bf 9}~~ No.\ {\bf 1},~~~ 5 -- 10}
   \setcounter{page}{5}

\author{Zhong-Lue Wen\inst{1},
        Jin-Lin Han\inst{1},
        Xiang-Yang Xu\inst{1,2},
        Yun-Ying Jiang\inst{1},
        Zhi-Qing Guo\inst{1},
        Peng-Fei Wang\inst{1}
        and
        Feng-Shan Liu\inst{1,3}}

\institute{National Astronomical Observatories, Chinese Academy of Sciences,
                 Beijing 100012, China;
                 {\it zhonglue@bao.ac.cn}, {\it hjl@bao.ac.cn}\\
\and
College of Science, Guizhou University, Guiyang 550025, China\\
\and
College of Physics Science and Technology,
                 Shenyang Normal University, Shenyang 110034, China}


\abstract{
We report the discovery of 4 strong gravitational lensing systems by
visual inspections of the Sloan Digital Sky Survey images of galaxy
clusters in Data Release 6 (SDSS DR6). Two of the four systems show Einstein
rings while the others show tangential giant arcs. These arcs or rings
have large angular separations ($>8''$) from the bright central galaxies
and show bluer color compared with the red cluster galaxies. In addition,
we found 5 {\it probable} and 4 {\it possible} lenses by galaxy clusters.
}

\authorrunning{Z. L. Wen  et al.}
\titlerunning{Discovery of four gravitational lensing systems by clusters in the SDSS DR6}

\noindent{\sf LETTERS}
\maketitle

\section{Introduction}

The strong gravitational lensing by galaxy clusters not only
constrains their mass distributions
\citep[e.g.][]{tc01,szb+08} but also provides an independent way to
study high-redshift faint background galaxies, which are otherwise unobservable
\citep[e.g.][]{mf93,mkm+03}.
Furthermore, the statistics of gravitational arcs in lensing clusters
can be used to constrain the paradigm of structure formation and
cosmology \citep[e.g.][]{bhc+98,lmj+05}.

Up to now, about 120 strong lensing clusters have been discovered
\citep[e.g.][]{lp89,sef+91,lgh+99,zg03,gky+03,ste+05,hgo+08}.
Almost all previous searches for the lensing systems are based on the
ground-based telescopes with seeing between $0.5''$ and $1.5''$.
\citet{ste+05} used the {\it HST} WFPC2 data and
found 104 giant arcs in 54 clusters.
Recently, based on 240 rich clusters identified from the Sloan
Digital Sky Survey \citep[SDSS,][]{yaa+00} and the follow-up deep
imaging observations from the Wisconsin-Indiana-Yale NOAO 3.5\,m
telescope and the University of Hawaii 88 inch telescope,
\citet{hgo+08} uncovered 16 new lensing clusters with giant arcs,
with 12 likely lensing clusters and 9 possible candidates.

In this letter, we report the discovery of 4 strong gravitational
lensing systems by clusters in the SDSS DR6.
The Einstein rings or giant arcs we found from the SDSS images suggest
that they are excellent strong lensing systems. In addition, we
found 5 {\it probable} and 4 {\it possible} lensing systems by galaxy
clusters.

\section{Lensing systems discovered from the SDSS DR6 images}

The SDSS provides five broad bands ($u$, $g$, $r$, $i$, and $z$) for
photometry and follow-up spectroscopy.  The star-galaxy separation is
reliable to the limit of $r=21.5$ \citep{lgi+01}. The photometric data
reaches the limit of $r=22.5$, with a mean seeing of 1.43$''$ in
$r$-band \citep{slb+02}.

Several valuable lenses have been {\it previously} discovered in the
SDSS data, though the survey is shallow (55~s exposure with a 2.5~m
telescope) making it difficult to detect faint giant arcs.
\citet{atl+07} have serendipitously found the ``8 O'clock arc'' from a
Lyman break galaxy at $z=2.73$ lensed by a luminous red galaxy SDSS
J002240.91+143110.4 at $z=0.38$.
\citet{bem+07,beh+09} looked for multiple blue objects around luminous
red galaxies, and discovered the ``Cosmic horseshoe'', an almost
complete Einstein ring of diameter $10''$ around SDSS
J114833.15+193003.5 at $z=0.444$, and also the ``Cheshire Cat'', which
consists of two galaxies at $z=0.97$ and $z>1.4$ lensed by a combination of
two giant early galaxies, SDSS J103847.95+484917.9 and SDSS
J103839.20+484920.3 at $z=0.426$ and 0.432, respectively. The lenses
in the above systems are galaxies.
Because there are a large number of high-redshift ($z\ge0.3$) clusters
in the SDSS, they can act as efficient lenses.  \citet{ead+07}
preformed a systematic search for giant arcs in 825 SDSS maxBCG
clusters \citep{kma+07} with masses $M\ge 2\times10^{14}~M_{\odot}$,
and found no gravitational arcs. However, they published a
serendipitous discovery of the Hall's arc lensed by a cluster, SDSS
J014656.0$-$092952 at $z=0.447$.
\citet{osk08} found the giant arc of a galaxy at $z=1.018$
lensed by a massive cluster SDSS J120923.7+264047 at $z=0.558$.
\citet{lba+08} found a galaxy at $z=2.00$ lensed by SDSS
J120602.09+514229.5, the brightest galaxy in a cluster.
\citet{sso+08} found a post-starburst galaxy at $z=0.766$ lensed by a
cluster galaxy at $z=0.349$.

%
{\it We searched for the lensing systems} by visual inspections of the
color images of a large sample of clusters.
We first searched for clusters by determining the luminous cluster
members ($M_r\le-21$) using the photometric redshifts of galaxies
brighter than $r=21.5$ \citep{cbc+03}. A cluster at $z$ is identified
when the number of member galaxies reaches 8 within a projected radius
of 0.5 Mpc and a photometric redshift ranges between $z\pm\Delta
z$. Here, we set $\Delta z=0.04(1+z)$ to allow variable
uncertainties of photometric redshifts at different redshifts. These
criteria can significantly reduce the false detection rate of clusters. From
the SDSS DR6, we identified $\sim40,000$ clusters of $0.05< z <0.6$
with an overdensity greater than 4.5 (Wen et al. in preparation).
Secondly, we searched for gravitational lensing features in these
clusters by inspecting the composite ($g$, $r$ and $i$) color images
from the SDSS web page\footnote{http://cas.sdss.org/astro/en/} independently by at
least by three authors.
We found 13 new candidates for lensing systems, in addition to the
known cases in 6 clusters, as listed in Table~\ref{lens.tab}.  In
addition to the name, redshift and the mass of cluster, we give
the angular separation of the arc from the bright central galaxy,
the magnitude and color of the brightest part of the arc and notes
in Table~\ref{lens.tab}.

\begin{table}[h!]

\centering

\begin{minipage}[]{100mm}

\caption[]{Lensing systems found to be associated with the SDSS
clusters.} \label{lens.tab}\end{minipage}

\vspace{-3mm} \fns


\tabcolsep 1.5mm
\begin{tabular}{lllllll}

\noalign{\smallskip}\hline\noalign{\smallskip}
Cluster name& $z$ & M$_{200}$ & $\theta$ & $r$ & $g-r$ & Reference, Notes\\
                      &                    & ({\tiny $10^{14}M_{\odot}$})& ($''$)   &     &       &\\
\hline\noalign{\smallskip}
SDSS J014656.0$-$092952 & 0.447 & $>$41.0& 13.6 & 21.13$\pm$0.09 & 0.54$\pm$0.13& 1, three giant arcs\\
NSC J082722.2+223244    & 0.335 &    23.8&  4.4 & 20.16$\pm$0.05 &--0.13$\pm$0.06& 2, multiple images\\
MACS J113313.1+500840   & 0.389 &    11.6& 10.0 & 21.62$\pm$0.13 &--0.10$\pm$0.16& 3, giant arc\\
SDSS J120602.0+514229   & 0.442 &  $>$7.7&  4.3 & 20.29$\pm$0.05 & 0.21$\pm$0.06& 4, arc near galaxy\\
SDSS J120923.6+264046   & 0.558 & $>$28.7& 11.3 & 22.37$\pm$0.26 & 0.82$\pm$0.42& 5, giant arc\\
NSCS J124034.0+450923   & 0.278 &    5.3 &  3.1 & 19.89$\pm$0.03 &--0.10$\pm$0.04& 6, arc near galaxy\\[1mm]
SDSS J090002.6+223404   & 0.489 &  $>$6.6&  8.4 & 20.41$\pm$0.06 & 0.06$\pm$0.07& 7*, almost certain:Einstein ring?\\
SDSS J111310.6+235639   & 0.324 &    26.8& 12.8 & 21.61$\pm$0.11 & 1.53$\pm$0.28& 7, almost certain:giant arc\\
SDSS J134332.8+415503   & 0.418 &    11.7& 12.4 & 21.09$\pm$0.12 & 0.10$\pm$0.16& 7, almost certain:giant arc\\
SDSS J223831.3+131955   & 0.413 &     9.7&  9.3 & 21.73$\pm$0.16 & 0.57$\pm$0.24& 7*, almost certain:Einstein ring\\[1mm]
SDSS J095739.1+050931   & 0.442 &  $>$5.8&  8.0 & 20.31$\pm$0.07 & 0.18$\pm$0.09& 7, probable:giant arc\\
SDSS J113740.0+493635   & 0.448 &  $>$5.6&  3.8 & 20.39$\pm$0.04 & 0.04$\pm$0.05& 6,7*, probable:arc near galaxy\\
maxBCG J120735.9+525459 & 0.278 &     7.0& 11.3 & 20.80$\pm$0.07 & 1.21$\pm$0.15& 7, probable:giant arc\\
SDSS J131811.5+394226   & 0.475 &  $>$6.6&  8.9 & 20.55$\pm$0.09 & 0.44$\pm$0.13& 7, probable:giant arc\\
NSCS J122648.0+215157   & 0.418 &    46.2& 10.8 & 21.74$\pm$0.19 & 0.17$\pm$0.22& 7, probable:giant arc\\[1mm]
SDSS J123736.2+553342   & 0.410 &    13.4&  4.6 & 20.14$\pm$0.04 & 0.18$\pm$0.05& 6,7*, possible:blue arc?\\
SDSS J131534.2+233301   & 0.517 &  $>$5.7&  5.5 & 20.36$\pm$0.08 & 0.96$\pm$0.15& 7, possible:arc?\\
SDSS J162132.3+060719   & 0.343 &     7.4& 16.2 & 19.88$\pm$0.06 & 1.20$\pm$0.13& 7, possible:giant arc?\\
SDSS J172336.1+341158   & 0.431 &  $>$8.5&  4.3 & 20.91$\pm$0.05 &--0.26$\pm$0.06& 7*, possible:multiple images?\\
\noalign{\smallskip}\hline\noalign{\smallskip}
\end{tabular}
\tablecomments{0.9\textwidth}{Here we list the name, redshift and
mass of cluster.
 The angular separation ($\theta$) between the arc
  and the central galaxy, the $r$-band magnitude and color ($g-r$) of the
  brightest part of arc, and also the references and notes on the
  lensing systems, are given in the following columns.  References:
  (1) \citet{ead+07}; (2) \citet{sso+08}; (3) \citet{ste+05}; (4)
  \citet{lba+08}; (5) \citet{osk08}; (6) \citet{beh+09}; (7) This
  work.  *: CASSOWARY candidates, see the end of Sect.3.}
\end{table}

The cluster mass, $M_{200}$, was estimated from the summed $r$-band
luminosity within $r_{200}$. Here, $r_{200}$ is the radius within
which the mean mass density is 200 times that of the critical cosmic
mass density. The mass-to-light ratio was calibrated by comparing the
cluster masses derived by \citet{rb02} with the summed $r$-band
luminosities of a sample of clusters (Wen et al. in preparation).
Such determined masses are the lower limits for clusters at $z>0.42$,
because their members are incomplete at the faint end of
$M_r=-21$.

\begin{figure}[h!!]

\vs\vs
\resizebox{35mm}{!}{\includegraphics{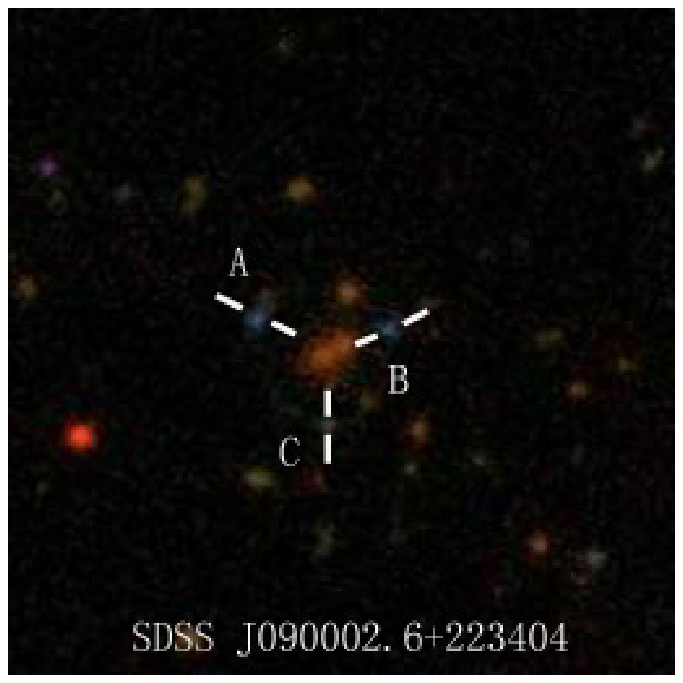}}~%
\resizebox{35mm}{!}{\includegraphics{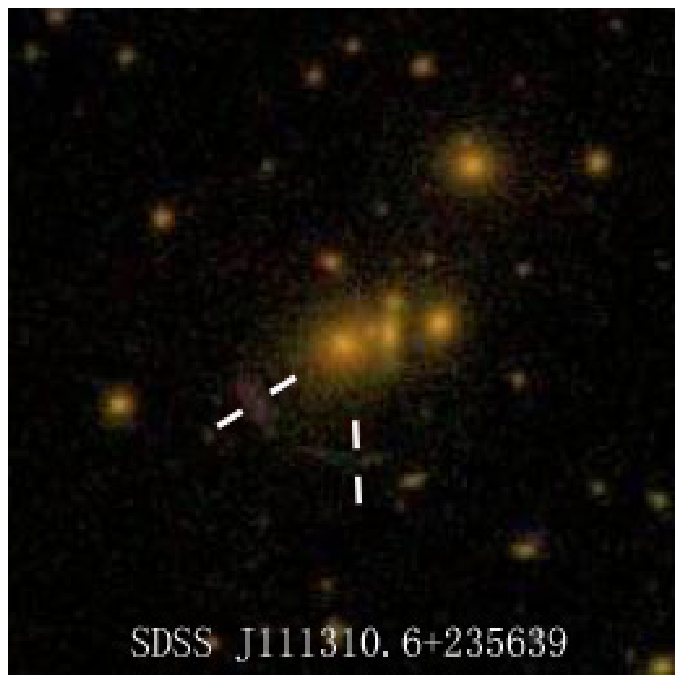}}~%
\resizebox{35mm}{!}{\includegraphics{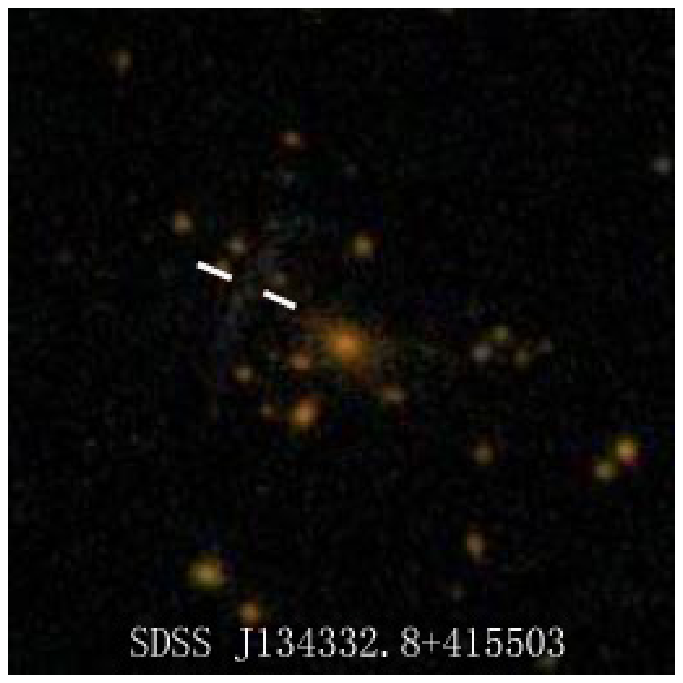}}~%
\resizebox{35mm}{!}{\includegraphics{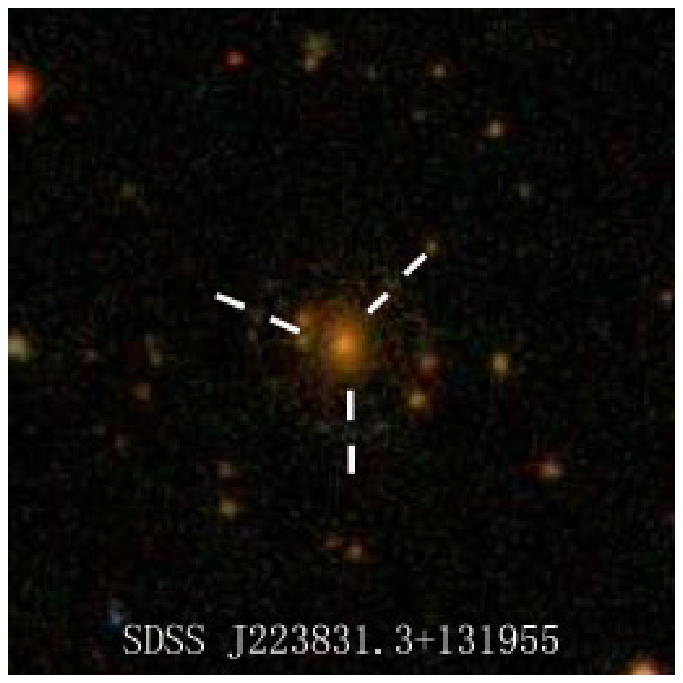}}\\[1mm]
\resizebox{35mm}{!}{\includegraphics{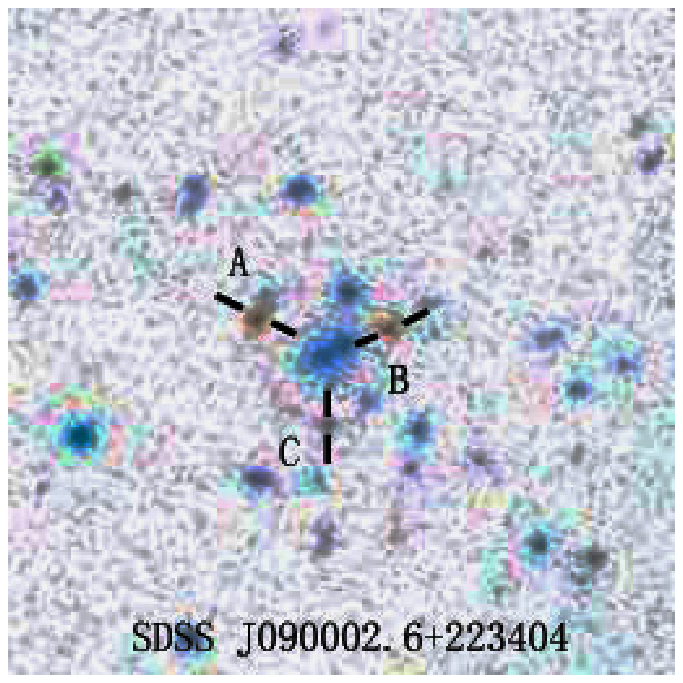}}~%
\resizebox{35mm}{!}{\includegraphics{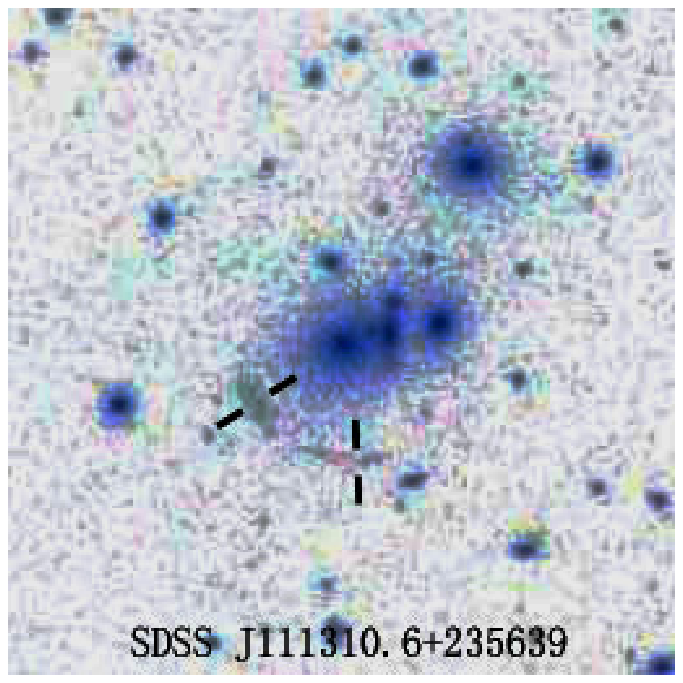}}~%
\resizebox{35mm}{!}{\includegraphics{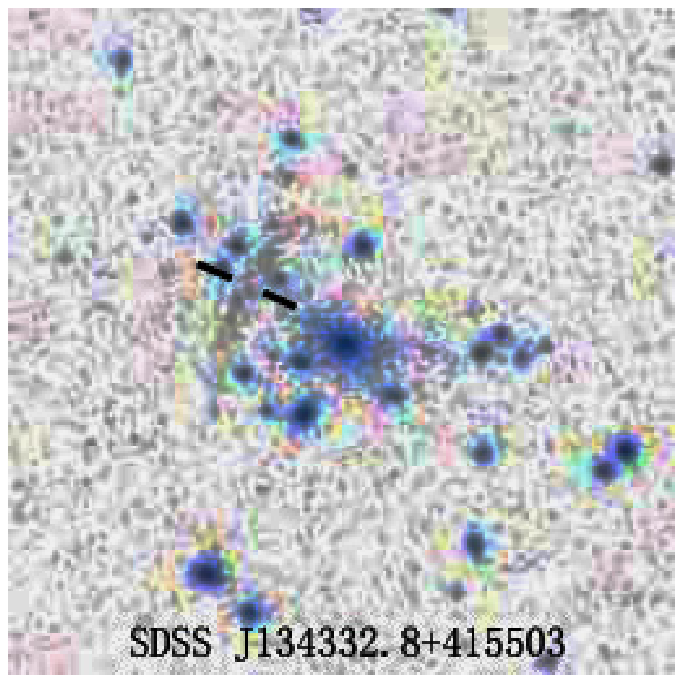}}~%
\resizebox{35mm}{!}{\includegraphics{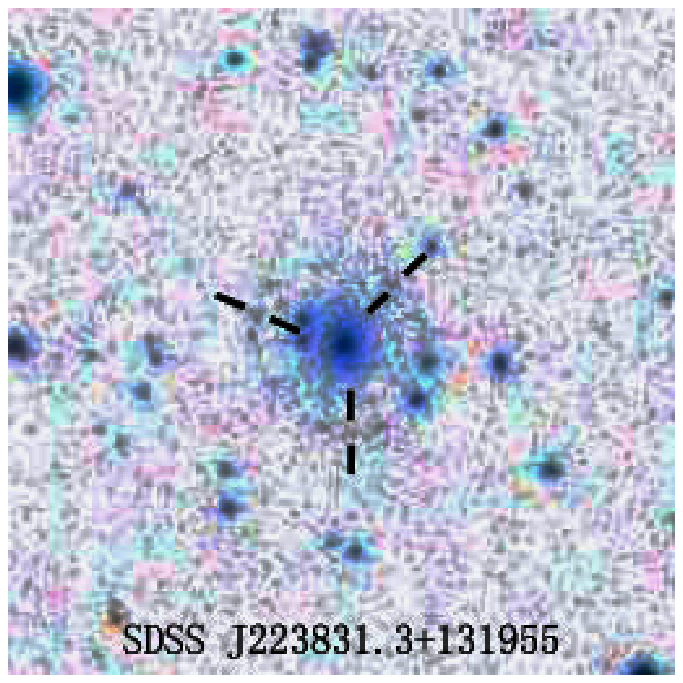}}\\[1mm]

\caption{\baselineskip 3.6mm
 The SDSS composite color images ($g$,
$r$ and $i$) of 4
  clusters which show Einstein rings or giant arcs. They
are almost certain gravitational lensing systems. The images
  have a field of view of 1.2$'\times$1.2$'$. The negative images are
  also shown in the second rows to see more clearly the lensing features.
\label{lens sure}}
\end{figure}

{\it By careful inspections of the SDSS images} of these
candidates, we almost certain found that 4 of them are lensing
systems (see Fig.~\ref{lens sure})   though the images are
somewhat shallow. For the system of SDSS J090002.6+223404, the
three arclets (A, B, C) around the two central galaxies nearly
form a circle with a radius of 8.4$''$, which may be an Einstein
ring if the arclets are the images of the same background source.
SDSS J223831.3+131955 shows a faint Einstein ring with a radius of
9.3$''$ . SDSS J111310.6+235639 and SDSS J134332.8+415503 show
tangential giant arcs with respect to the bright central galaxies
with a separation of more than $12''$. The faint but clear arcs
usually have blue color compared with the bright central galaxies.

\begin{figure}
\resizebox{35mm}{!}{\includegraphics{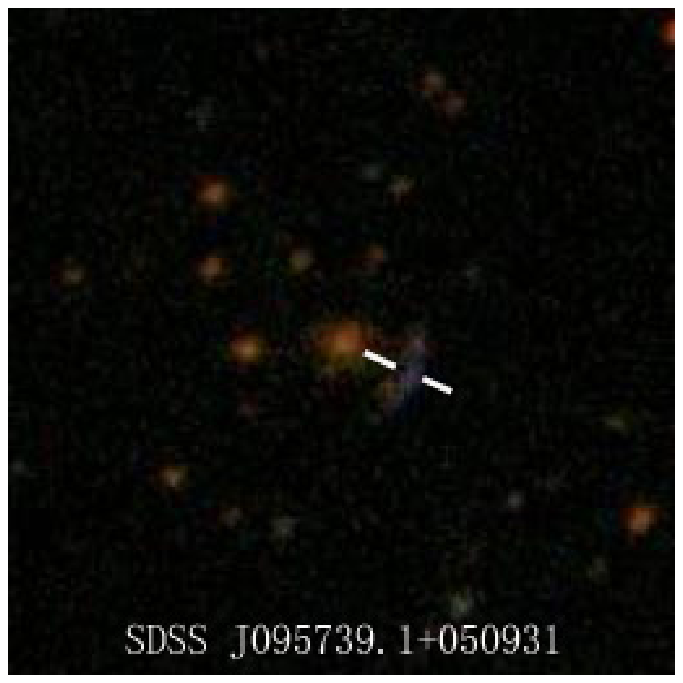}}~%
\resizebox{35mm}{!}{\includegraphics{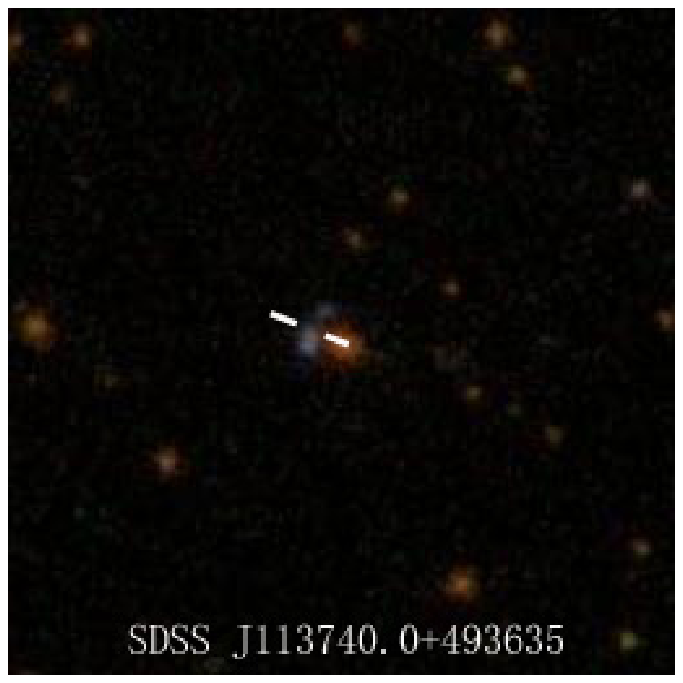}}~%
\resizebox{35mm}{!}{\includegraphics{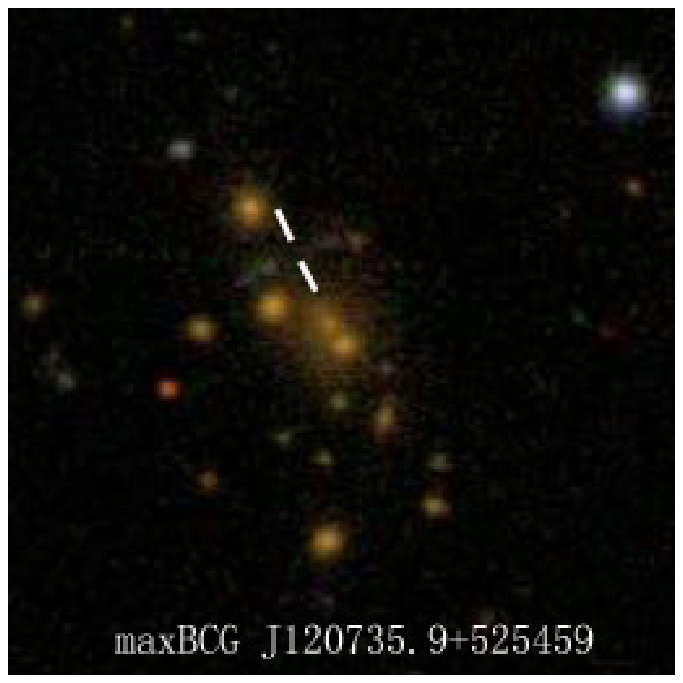}}~%
\resizebox{35mm}{!}{\includegraphics{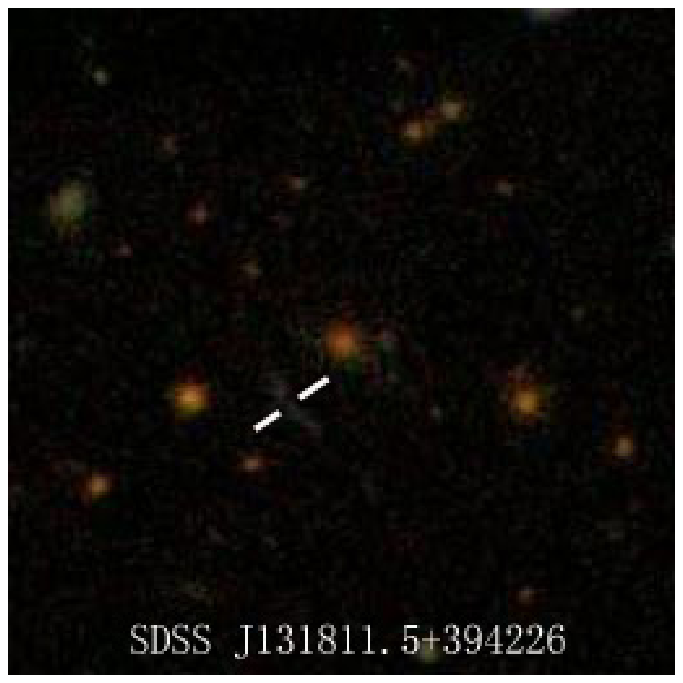}}\\[1mm]
\resizebox{35mm}{!}{\includegraphics{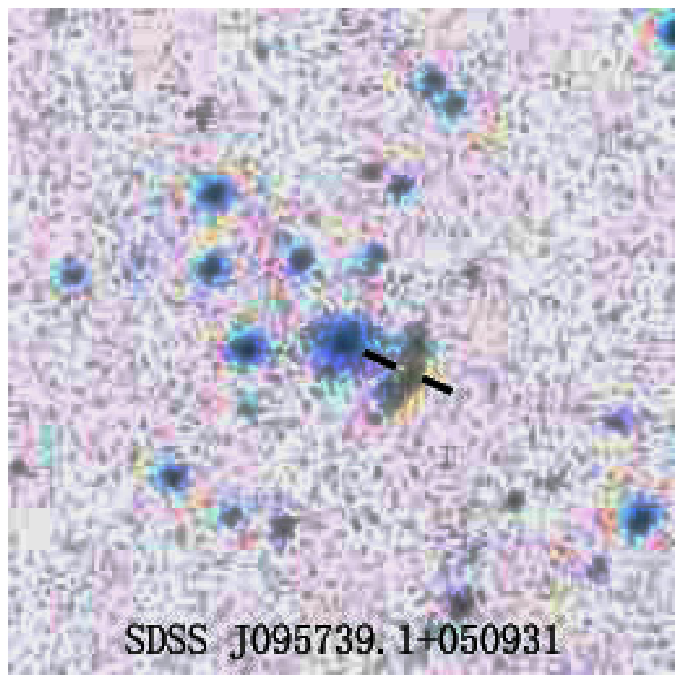}}~%
\resizebox{35mm}{!}{\includegraphics{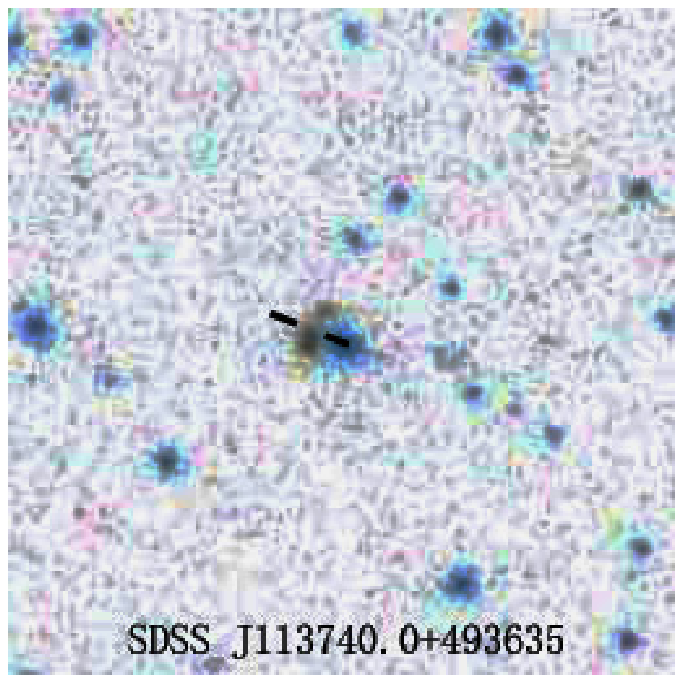}}~%
\resizebox{35mm}{!}{\includegraphics{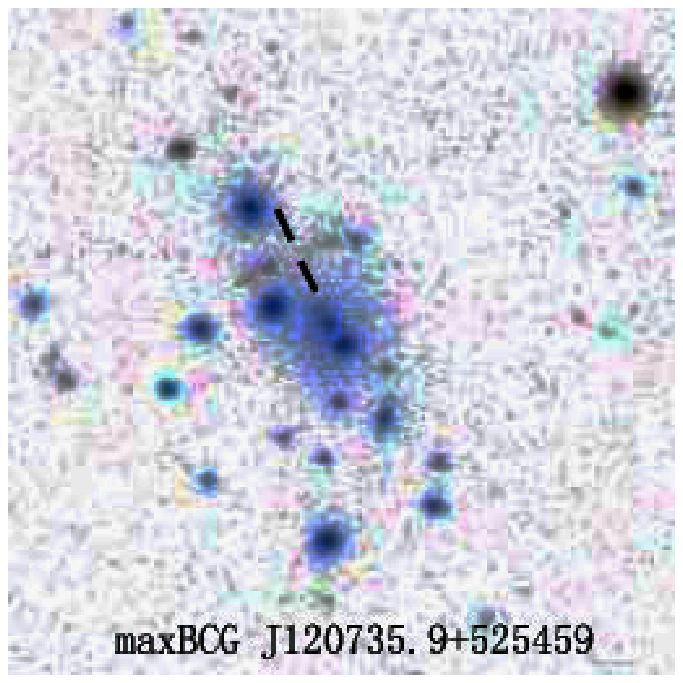}}~%
\resizebox{35mm}{!}{\includegraphics{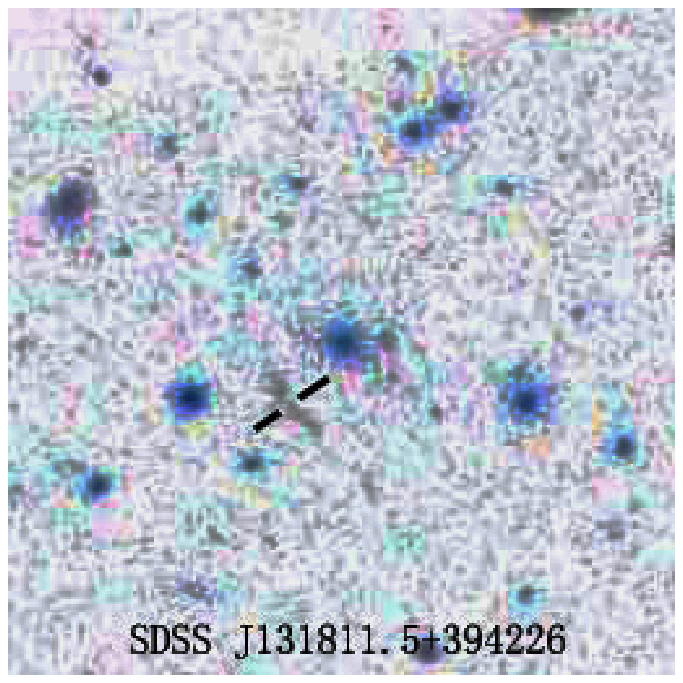}}\\[1mm]
\resizebox{35mm}{!}{\includegraphics{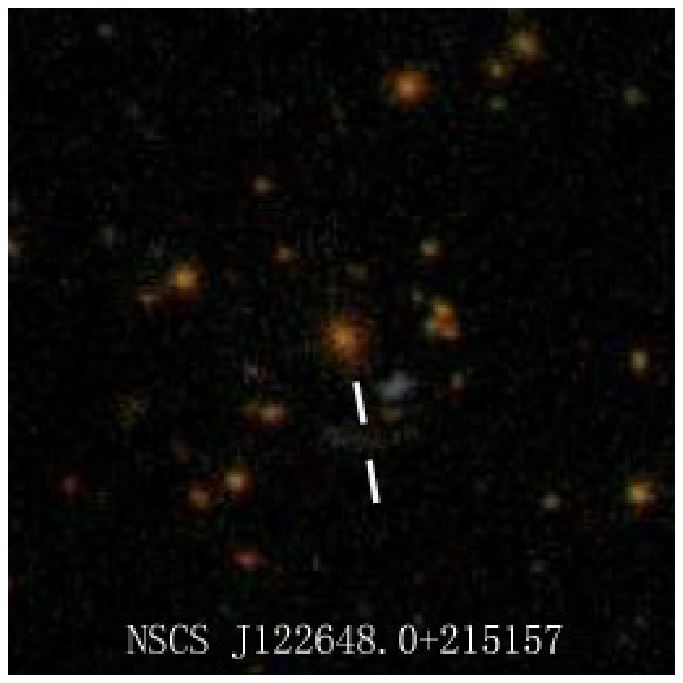}}~%
\resizebox{35mm}{!}{\includegraphics{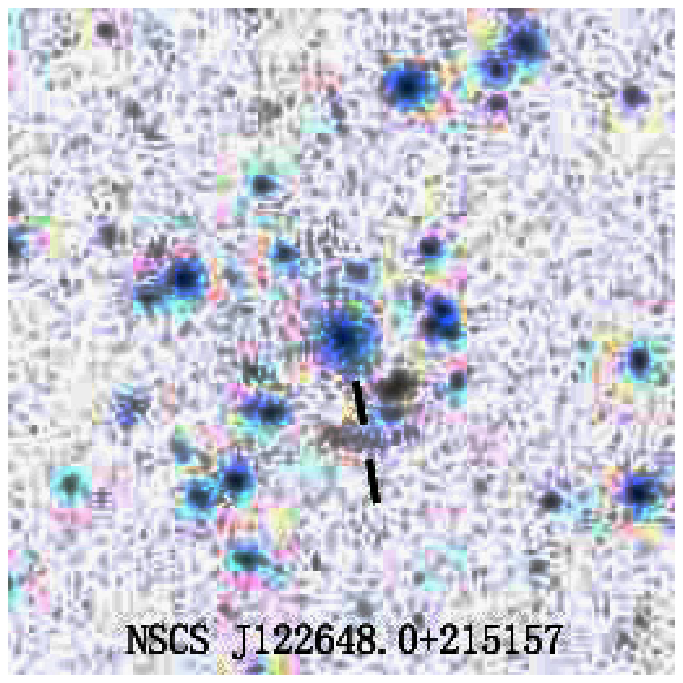}}

\caption{\baselineskip 3.6mm
Same as Fig.~\ref{lens sure}, but for
5 clusters which are {\it probable} lensing systems. The system of
NSCS J122648.0+215157 in the last row was found from a merging
cluster during the proof-reading stage of this paper.
\label{lens prob}}
%
\resizebox{35mm}{!}{\includegraphics{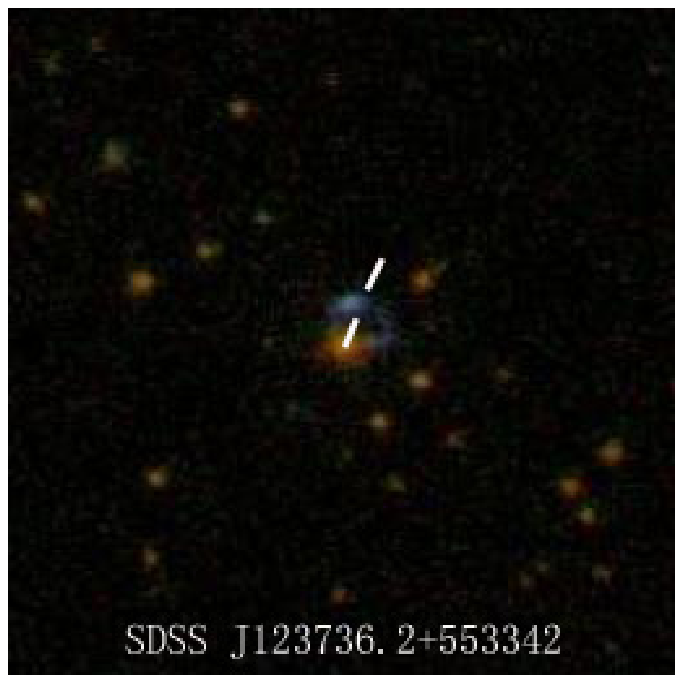}}~%
\resizebox{35mm}{!}{\includegraphics{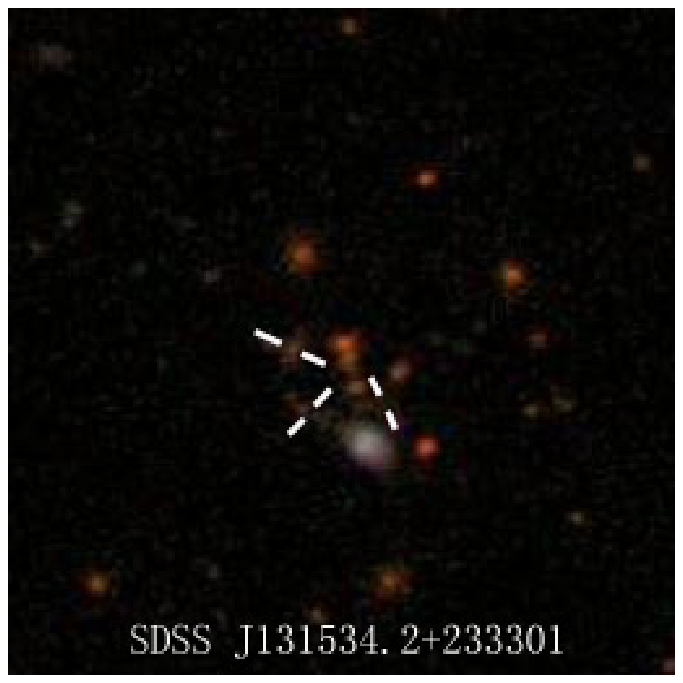}}~%
\resizebox{35mm}{!}{\includegraphics{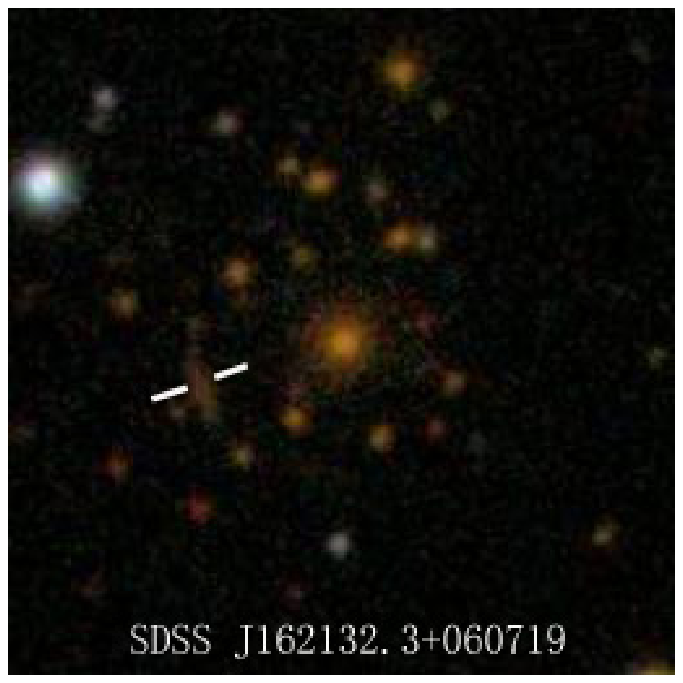}}~%
\resizebox{35mm}{!}{\includegraphics{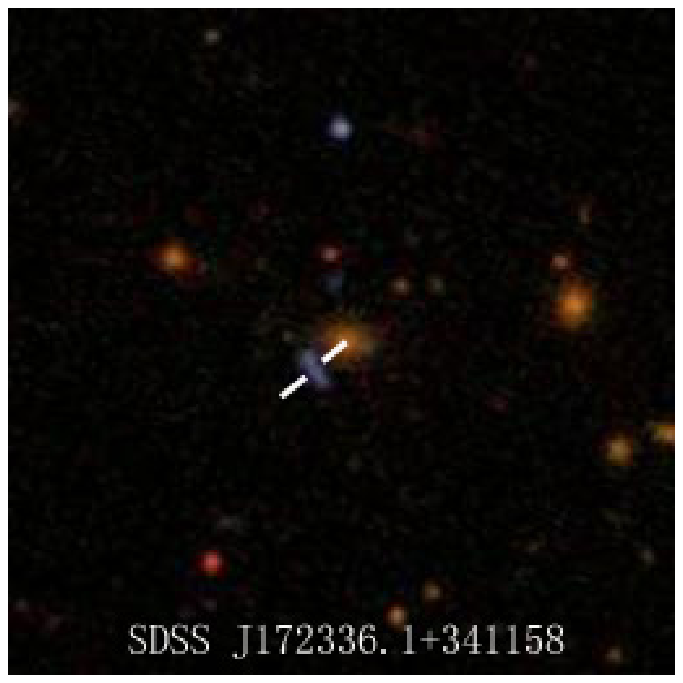}}\\[1mm]
\resizebox{35mm}{!}{\includegraphics{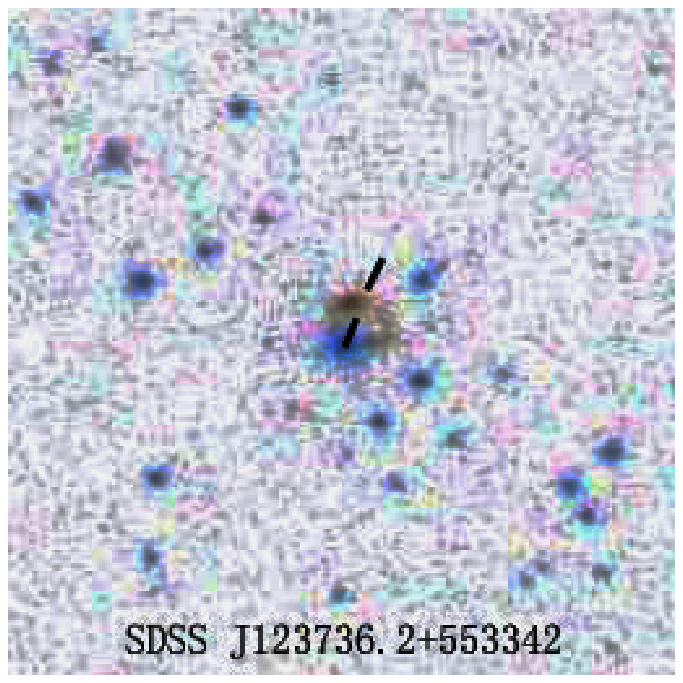}}~%
\resizebox{35mm}{!}{\includegraphics{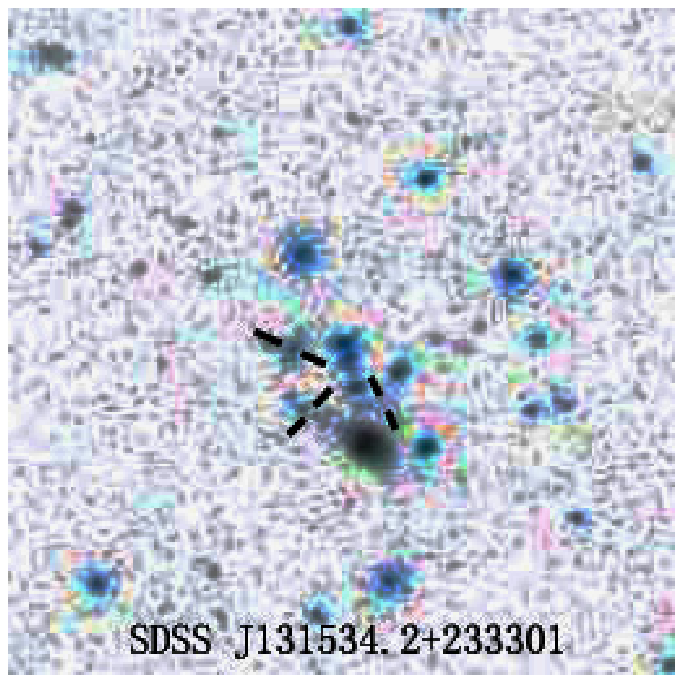}}~%
\resizebox{35mm}{!}{\includegraphics{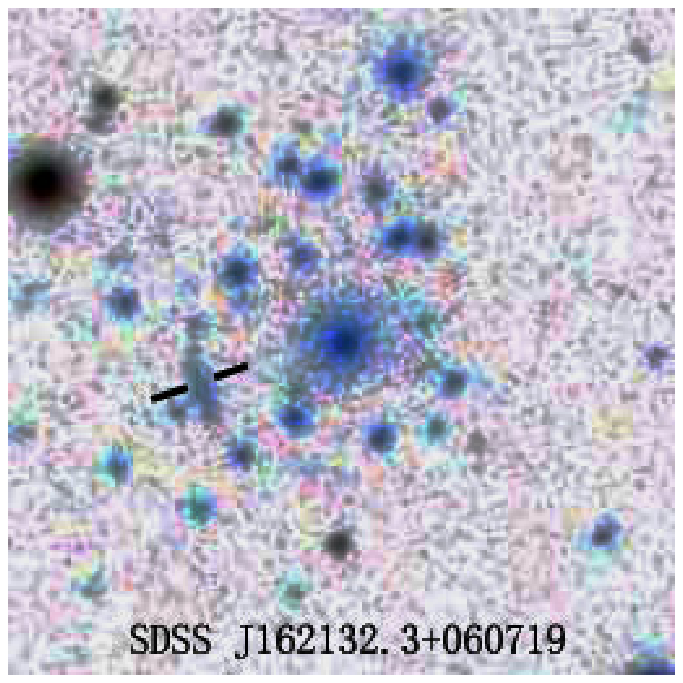}}~%
\resizebox{35mm}{!}{\includegraphics{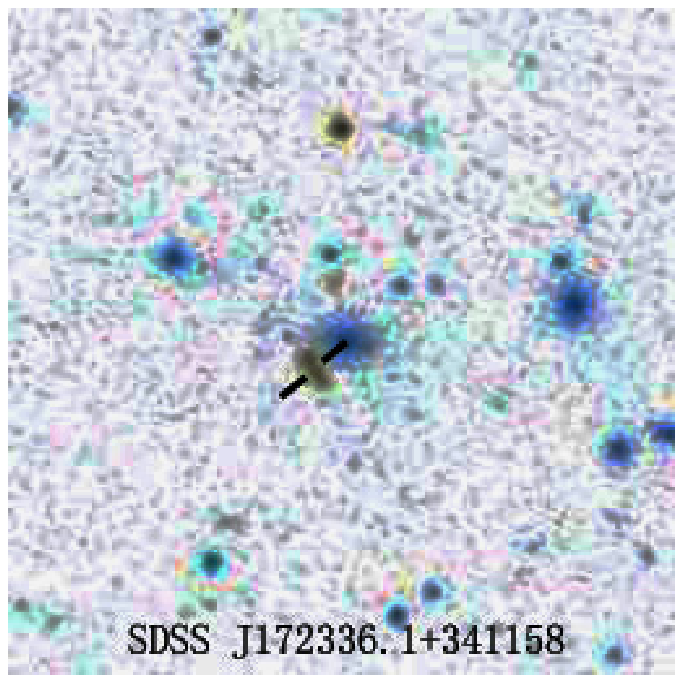}}\\[1mm]

\centering
\begin{minipage}[]{105mm}

\caption{Same as Fig.~\ref{lens sure}, but for 4 clusters which
are {\it possible} lensing systems.
\label{lens poss}}\end{minipage}
\end{figure}

The total mass (including dark matter) within the
Einstein ring $r_{\rm E}=D_{\rm l} \theta_{\rm E}$ can be estimated by,
\begin{equation}
M(<r_{\rm E})=\frac{c^2r_{\rm E}^2}{4G}\frac{D_{\rm s}}{D_{\rm
    l}D_{\rm ls}},
\end{equation}
where $D_{\rm s}$ and $D_{\rm l}$ are the angular diameter distances of
the source and lens from the observer, and $D_{\rm ls}$ is the angular
diameter distance of the source from the lens. We estimate the angular
Einstein ring $\theta_{\rm E}$ by $\theta$; this is an approximation due to
the non-sphericity of clusters. In an $\Lambda$CDM
cosmology (H$_0=$72 ${\rm km~s}^{-1}$ ${\rm Mpc}^{-1}$, $\Omega_m=0.3$
and $\Omega_{\Lambda}=0.7$), we obtained the mass $M(<r_{\rm
  E})=2.3\times10^{13}~M_{\odot}$ for SDSS J090002.6+223404 if we
assume the source redshift of $z_{\rm s}=1$, or $M(<r_{\rm
  E})=1.6\times10^{13}~M_{\odot}$ if $z_{\rm s}=2$.  Similarly, for
SDSS J223831.3+131955, the mass is $M(<r_{\rm
  E})=2.3\times10^{13}~M_{\odot}$ if $z_{\rm s}=1$ or $M(<r_{\rm
  E})=1.7\times10^{13}~M_{\odot}$ if $z_{\rm s}=2$.

We also found another 5 clusters which are {\it probable} lensing
systems (see Fig.~\ref{lens prob}). All of them show blue arclets,
which are tangential to the bright central galaxies and are distinct
in color from the red cluster galaxies. Notably,
the giant arcs in the clusters, SDSS J095739.1+050931 and 
NSCS J122648.0+215157, are faint,
and have very large separations ($>10''$) from the central galaxy.
SDSS J113740.0+493635 has a blue arc, which is very close ($3.8''$) to
the central red galaxy.
In the cluster SDSS maxBCG J120735.9+525459, the blue giant arc stands
out among the red member galaxies.

Another 4 clusters are {\it possible} lensing systems (see
Fig.~\ref{lens poss}).
The blue images around the central galaxy of SDSS J123736.2+553342 may
be the lensed image of a background source. However, this bright blue
object on the top can also be a foreground galaxy.
In the cluster SDSS J131534.2+233301, the three faint arclets may form
a half ring surrounding several member galaxies rather than the
brightest galaxy. These arclets may be independent, or they may just be
faint member galaxies.
In the cluster SDSS J162132.3+060719, the arc is tangential to the
bright central galaxy and has a large separation ($16.2''$), but does not have
much color difference from the cluster galaxies. It may also be
due to a combination of an edge-on galaxy plus other faint objects.
The outstanding blue arclet in SDSS J172336.1+341158 is very close ($4.3''$) to
the central red galaxy. It may be the lensed and enhanced image of the
background galaxy, but the possibility of the foreground galaxy can
not be excluded.

\section{Final Remarks}

Using the SDSS data, we almost certainly found 4 cluster lenses, plus 5
probable and 4 possible lenses. Together with 6 known lenses of
clusters, there are at least 10 lensing systems identified from SDSS, which
preferably have redshifts around 0.4 and masses of at least
$5\times10^{14}~M_{\odot}$. The separations of the blue arcs or rings from
the central red galaxy are usually several arcseconds.

If the 19 clusters listed in Table~\ref{lens.tab} are all lenses, 6 of
them have redshifts of $0.2<z<0.4$, and 13 of $0.4<z<0.6$.
Comparing them with 7568 and 10121 clusters with masses
$5\times10^{14}~M_{\odot}$ in the corresponding redshift ranges (Wen
et al. in preparation), we found that the occurrence probability of
lensing clusters in the shallow SDSS images increases from
$7.9\times10^{-4}$ in the range of $0.2<z<0.4$ to $1.3\times10^{-3}$
of $0.4<z<0.6$. The tendency of the increase in lensing probability with redshift
is consistent with that of \citet{gky+03}.

Follow-up observations are necessary to confirm the {\it probable} or {\it possible}
lensing systems in 9 clusters. Unfortunately, we can not easily access
large optical telescopes to make the follow-up confirmation.
We also publish these candidates with a list and the SDSS color
images to encourage followup observations by others. After we submitted this paper
to this journal and to astro-ph, we learned from Dr.~V.~Belokurov that 5
objects with ``*'' in Table~\ref{lens.tab} have been listed in their
CASSOWARY catalogue (see http://www.ast.cam.ac.uk/research/cassowary/) as lensing candidates
CSWA 19, CSWA 10, CSWA 7, CSWA 13 and CSWA 14, respectively. The
images of CSWA 7 and CSWA 13 have been published in
\citet{beh+09}. These 5 objects therefore are independent ``re-discoveries''.

\noindent{\bf NOTES ADDED IN PROOF:} 1) We found another probable lensing system
by a merging cluster, NSCS J122648.0+215157 \citep{ldg+04}, which we have added to
Fig.~\ref{lens prob} and Table~\ref{lens.tab}; 2) We noted that Kubo
et al. (2008) just submitted a paper to report their SDSS arc
survey results, including the follow-up observations of two lensing
systems we found independently, SDSS J111310.6+235639 and
J113740.0+493635.

\normalem
\begin{acknowledgements}

We thank Prof. Xiang-Ping Wu and Shude Mao for a carefully reading of the
manuscript and the referee for helpful comments.
The authors are supported by the National Natural Science
Foundation  of China (NSFC, Nos.10521001, 10773016 and 10833003)
and the National Key Basic Research Science Foundation of China
(2007CB815403).
Funding for the SDSS and SDSS-II has been provided by the Alfred
P. Sloan Foundation, the Participating Institutions, the National
Science Foundation, the U.S. Department of Energy, the National
Aeronautics and Space Administration, the Japanese Monbukagakusho, the
Max Planck Society, and the Higher Education Funding Council for
England.

\end{acknowledgements}

\bibliographystyle{raa}
\bibliography{journals,lens}

\end{document}